\documentclass[11pt]{article}
\usepackage{fullpage}

\usepackage{xspace} 
\usepackage{alltt}

\usepackage[T1]{fontenc}

\usepackage{hyperref}

\usepackage{xcolor}
\definecolor{codegreen}{rgb}{0,0.6,0}
\definecolor{codegray}{rgb}{0.5,0.5,0.5}
\definecolor{codepurple}{rgb}{0.58,0,0.82}
\usepackage{listings}
\makeatletter
\lst@AddToHook{OnEmptyLine}{\addtocounter{lstnumber}{-1}}%
\makeatother
\lstdefinestyle{pystyle}{
    backgroundcolor=\color{white},
    basicstyle=\footnotesize\ttfamily, %
    morecomment = [is]{\#\#\#}{\#\#},      %
    commentstyle=\color{brown},
    keywordstyle=\color{blue}, %
    stringstyle=\color{codepurple},
    numberstyle=\tiny\color{codegray},
    breakatwhitespace=false,
    breaklines=false,
    captionpos=b,
    keepspaces=true,                 
    numbers=left,
    numbersep=5pt,
    numberblanklines=false,
    showspaces=false,                
    showstringspaces=false,
    showtabs=false,                  
    tabsize=2
}
\makeatother
\def\progdir{polling} %

\newcommand{\todo}[1]{} %
\newcommand{\notes}[1]{} %

\newcommand{\Hex}[1]{\hspace{#1ex}}
\newcommand{\Vex}[1]{\vspace{#1ex}}

\newenvironment{code}{\Vex{0.5}\begin{alltt}\small}{\end{alltt}\Vex{0.5}}
\newcommand\co[1]{\mbox{\tt\small #1}} %
\newcommand\p[1]{\m{#1}}
\newcommand\m[1]{\mbox{$#1$}} %
\newcommand{\defn}[1]{#1} %
\def\mathify#1{\ifmmode{\mbox{$#1$}}\else\mbox{$#1$}\fi}

\newcommand{\mypar}[1]{\Vex{1.5} \noindent {\bf #1.~}}

\begin{document}

\newcommand\ack{
This work was supported in part by NSF under grants 
  CCF-1414078, %
  CCF-1954837, %
  CNS-1421893, %
  and IIS-1447549, %
  and ONR under grant N00014-20-1-2751. %
}

\title{Assurance~of~Distributed~Algorithms~and~Systems:\\
  Runtime Checking of Safety and Liveness\thanks{\ack}}

\author{Yanhong A. Liu \Hex{10} Scott D. Stoller\\
Computer Science Department, 
  Stony Brook University, Stony Brook, NY, USA\\
  \{liu,stoller\}@cs.stonybrook.edu}

\date{}

\maketitle

\begin{abstract}
  This paper presents a general framework and methods for complete
  programming and checking of distributed algorithms at a high-level, as in
  pseudocode languages, but precisely specified and directly executable, as
  in formal specification languages and practical programming languages,
  respectively.
  The checking framework, as well as the writing of distributed algorithms
  and specification of their safety and liveness properties, use DistAlgo,
  a high-level language for distributed algorithms.
  We give a complete executable specification of the checking framework,
  with a complete example algorithm and example safety and liveness
  properties.
\end{abstract}

\section{Introduction}
\label{sec-intro}

Distributed systems are increasingly important in our increasingly
connected world.  Whether for distributed control and coordination or
distributed data storage and processing, 
at the core are distributed algorithms.

It is well known that distributed algorithms are difficult to understand.
That has led to significant effort in specifying these algorithms and
verifying their properties,
e.g.,~\cite{hawblitzel2015ironfleet,Cha+16PaxosTLAPS-FM,padon2017paxos},
as well as in developing specification languages and verification tools, e.g.,
TLA and TLA+ Toolbox~\cite{lam94tla,lam02book,tlatoolbox20}, I/O
Automata~\cite{Lynch96}, and Ivy~\cite{mcmillan2020ivy}.  However,
challenges remain for automated verification of practical distributed
algorithms using theorem proving or model checking techniques, due to
exorbitant manual effort and expertise required or prohibitive state-space
explosion.

Runtime verification allows information to be extracted from a running
system and used to check whether observed system behaviors satisfy certain
properties, and to react based on the results of checking.  It is the most
effective complement to theorem proving and model checking %
for sophisticated algorithms and implementations.  For routinely checking
real distributed applications written in general-purpose programming
languages, it is so far the only feasible practical solution.

Many methods and related issues for doing such runtime checking have been
studied, as discussed in Section~\ref{sec-related}.  Such checking and all
kinds of variations have also long been used extensively in practical
program development, testing, debugging, and simulation for distributed
algorithms.  However, these studies and uses of runtime checking are
either more abstract methods, not presented as executable programs, or
involving significant programming using commonly-used programming languages,
too large to present in complete and exact forms on paper.  

This paper presents a general framework and methods for complete
programming and checking of distributed algorithms at a high-level, as in
pseudocode languages, but precisely specified and directly executable, as
in formal specification languages and practical programming languages,
respectively.
The checking framework, as well as the writing of distributed algorithms
and specification of their safety and liveness properties, use DistAlgo, a
high-level language for distributed
algorithms~\cite{Liu+17DistPL-TOPLAS,distalgo20git}.
We give a complete executable specification of the checking framework, with a
complete example algorithm and example safety and liveness properties.

The framework can check any desired properties
against observed system behavior.  
Note that since any execution of a practical system is finite, the liveness
properties we check are bounded liveness, that is, the desired properties
hold within specified time bounds.
The framework requires no change to the algorithm code to be checked.  It
puts the algorithm code, property specification, as well as fault
simulation together with small configurations, thanks to the power of the
DistAlgo language and compiler.  The complete checking program then
automatically intercepts messages sent and received by the distributed
processes to be checked, with both logical and real times, and checks the
specified properties at desired points as written.

This framework has been used in implementing, testing, debugging,
simulation, and analysis of many well-known distributed algorithms, and in
teaching.  Our experiences included discovering improvements to both
correctness and efficiency of some well-known algorithms,
e.g.,~\cite{Liu+12DistSpec-SSS,Liu18LclockUnfair-APPLIED,Liu+19Paxos-PPDP}.

\section{Distributed algorithms and their safety and liveness}
\label{sec-dist}

Distributed algorithms are algorithms that run in distributed systems.
Understanding and verifying their properties are central challenges for
distributed computing.

\mypar{Distributed systems and distributed algorithms}
A distributed system is a set of distributed processes.  Each process has
its own private memory that only it can access.  Processes
execute concurrently and communicate with each other by sending and
receiving messages.

Distributed processes and communications are prone to various kinds of
failures, depending on the underlying infrastructures.  Processes may be
slow, may crash, may later recover, and may even behave arbitrarily.
Messages may be lost, delayed, duplicated, reordered, and even be
arbitrarily changed.

Distributed algorithms are for solving problems that involve coordination,
negotiation, etc.\ among distributed processes in the presence of possible
failures.
Due to nondeterminism from concurrency and uncertainty from failures,
distributed algorithms are notoriously difficult to design, understand, and
verify. %
Even as an algorithm executes in a distributed system, the state of the
system is not fully observable, and the order of events cannot be fully
determined.  This led to Lamport's creation of logical clocks, which are
fundamental in distributed systems~\cite{Lam78}.

Distributed computing problems are of an extremely wide variety, and a 
great number of distributed algorithms have been studied.
e.g.,~\cite{Lynch96,Garg02,Fok13}.
Well-known problems range from distributed clock synchronization to
distributed snapshot, from leader election to distributed mutual exclusion,
from atomic commit to distributed consensus, and many more.  We give two
examples here:
\begin{itemize}

\item Distributed mutual exclusion.  Distributed mutual exclusion is for
  multiple processes to access a shared resource mutually exclusively,  
  in what is called a critical section, i.e., there can be at most one
  process in a critical section at a time.

  It is one of the most studied problems,
  e.g.,~\cite{raynal1986algorithms,KshSin08}, with at least dozens if not
  hundreds or more of proposed algorithms and variants.
  For example, Lamport's algorithm~\cite{Lam78}, introduced to show the
  use of logical clocks, was designed to guarantee that access to the resource
  is granted in the order
  of logical clock values of the requests.

\item Distributed consensus.  Distributed consensus is for a set of
  processes to agree on a single value or a continuing sequence
  of values, called \defn{single-value consensus} or \defn{multi-value
    consensus}, respectively.

  It is essential in any important service that maintains a state,
  including services provided by companies like Google and Amazon.  This is
  because such services must use replication to tolerate failures caused by
  machine crashes, network outages, etc.  Replicated processes must agree on
  the state of the service or the sequence of operations that have been
  performed, e.g., that a customer order has been placed and paid but not yet
  shipped, so that when some processes become unavailable, the remaining
  processes can continue to provide the service correctly. 

  Even well-known algorithms and variants number at least dozens, 
  starting from virtual
  synchrony~\cite{birman1987reliable,birman1987vs,Bir+10virtualsync},
  viewstamped replication~\cite{oki88vsr,liskov2012vr}, and
  Paxos~\cite{Lam98paxos}.
\end{itemize}
These problems are at the core of distributed file systems, distributed
databases, and fault-tolerant distributed services in general.  New
algorithms and variants for them are developed constantly, not to mention a
vast number of other distributed algorithms, such as network protocols,
distributed graph algorithms, and security protocols.

\mypar{Safety and liveness}
Lamport~\cite{lamport1977proving} first formulated two types of properties
of a distributed system: safety and liveness.  Informally, a safety property states
that some bad things will not happen, and a liveness property states that some
good things will happen.
We continue the two examples discussed earlier:
\begin{itemize}

\item For distributed mutual exclusion, a most important safety
  property is that at most one process is in a critical session at a
  time.
  A liveness property is that some requests are eventually served, and a
  stronger liveness property is that all requests are eventually served.

  For example, Lamport's algorithm~\cite{Lam78} is designed to guarantee all
  these, and in fact, as mentioned above, to guarantee a stronger
  property---that all requests are served in the order of logical clock
  values.  This stronger property can be interpreted and formulated
  as either a safety property, to mean that no requests are served out of
  the order of logical clock values, or a liveness property, to include
  that all requests are eventually served.

\item For distributed consensus, there are two important safety
  properties: (1) agreement on the decided single value, in single-value
  consensus, or on the sequence of values, in multi-value consensus, by
  nonfaulty processes, and (2) validity of the decided value or values to
  be among allowed values.  A liveness property for single-value consensus
  is that nonfaulty processes eventually decide on a value.   A liveness property
  for multi-value consensus is that nonfaulty processes repeatedly decide on 
  additional values in the sequence.

  Good distributed consensus algorithms, such as Paxos~\cite{Lam98paxos},
  guarantee the two safety properties, but they cannot guarantee the
  liveness property
  due to the well-known impossibility of consensus in asynchronous distributed 
  systems even with only one faulty process subject to crash failures~\cite{fischer85flp}.
\end{itemize}
Specifying safety and liveness properties is nontrivial, especially
liveness properties, even informally.  For example, liveness for many
consensus algorithms and variants has been left unspecified, or specified
too weakly to be useful or too strongly to be possible~\cite{ChaLiu20Liveness-arxiv}.

Safety and liveness are, in general, global properties about multiple processes.
Checking them requires knowing the states of multiple processes.
However, the state of a process is private to that process and cannot be
accessed directly by any other process.
The best one can do is to observe a process by intercepting messages sent
and received by that process, and determine the state of the system and
desired properties conservatively or approximately, and with a delay.

We use \defn{checker} to refer to a process that observes the sending and
receiving of messages by a set of processes and checks desired properties.

\section{A powerful language for distributed programming}
\label{sec-lang}

A powerful language for distributed programming must allow (1) easy
creation of distributed processes and communication channels for sending
messages, (2) easy handling of received messages, both synchronously (with
waiting) and asynchronously (without waiting), (3) easy processing of all
information communicated as well as a process's own data, and (4) easy
configuration of basic elements for real execution on distributed machines.

\mypar{Running example: The polling problem}
We introduce a simple but essential problem, which we call the polling
problem, as a running example: 
\begin{quote}
  A poller process sends a question to a set of pollee processes, 
  waits to receive answers to the question from all of them, and then sends 
  an outcome message to them.
\end{quote}
Small variations of this problem include waiting to receive replies from a
subset of the pollee processes, such as a majority or a quorum,
instead of all of them.

This problem is essential because any process working with a set of other
processes requires talking to and hearing back from those processes one way
or another.  This problem manifests widely in well-known distributed
algorithms, including algorithms for distributed mutual exclusion, atomic
commit, and distributed consensus.  This problem also manifests itself in
everyday life, such as an instructor giving assignments to students, a
chairperson soliciting suggestions from committee members, or a campaign
organizer sending a poll to voters.

The problem appears simple, but is nontrivial, even without process
failures or message delays or losses, because processes are generally
communicating with multiple processes and doing other things at the same
time.  Consider some examples:
\begin{itemize}

\item When the poller receives a message from a pollee, how does the
  poller know it is replying to a particular question?  The pollee might
  happen to send something to the poller with the same format as an
  expected reply, and send it shortly after the question was sent.

\item How does the poller know it has received replies from all pollees?
  It could compare the number of replies to the number of pollees, but a
  pollee might send multiple replies, or a communication channel might
  duplicate messages.

\end{itemize}
The problem becomes even harder if processes can fail and messages may be
lost or delayed.  It becomes even more challenging if processes can fake
identities and messages can be altered or counterfeited.  In the latter
cases, processes need to use security protocols for authentication and
secure communication.  Although we do not consider those problems further
in this tutorial, we have extended DistAlgo with a high-level cryptographic
API for expressing such security protocols~\cite{Kan+18CryptoAbs-PLAS}.

\begin{figure}[ht!]
  \centering
\begin{lstlisting}
class P (process):            # define poller process class
  def setup(rs): pass         # take a set of responder processes
  def run(): ###
#    polling()
#    send('done', to=parent()) # send done to main
#
#  def polling(): ##
    self.t = logical_clock()  # get logical clock time as question id
    send(('question', question(), t), to= rs)
    await each(r in rs, has= some(received(('reply',_,_t), from_=_r)))
    output('-- received Y from:', setof(r, received(('reply','Y',_t), from_= r)))
    send(('outcome', outcome()), to= rs)

  def question(): return 'Did you attend RV (Y/N)?' # create question
  def outcome(): return countof(r, received(('reply','Y',_t), from_= r))

class R (process):            # define responder (i.e., pollee) process class
  def run(): ###
#    polled()
#
#  def polled(): ##
    await some(received(('outcome', o)))
    output('== received outcome:', o)
  def receive(msg=('question', q, t), from_= p):
    send(('reply', reply(q), t), to= p)

  def reply(q): import random; return random.choice(['Y','N']) # create reply

def main():
  config(clock = lamport)     # use Lamport clock
  config(channel = reliable)  # use reliable channels
  rs = new(R, [], num= 10)    # create and set up 10 responders
  p = new(P, [rs])            # create and set up poller process p
  start(rs)                   # start responders
  start(p)                    # start poller
###
#  await received('done', from_= p) # add at end of file
#  end(rs)                     # end pollers ##
\end{lstlisting}
  \caption{Polling program in DistAlgo.}
  \label{fig-polling}
\end{figure}

Figure~\ref{fig-polling} shows a complete polling program written in
DistAlgo.  Process classes \co{P} and \co{R} specify the poller and
responder (i.e., pollee) processes, respectively.  Definitions \co{run} and
\co{receive} specify the main algorithm.
The core of the algorithm is on lines 4-6, 8, 13, and 15-16.
The rest puts all together, plus setting up processes, starting them, and
outputting about replies and outcomes of the polling.  The details are
explained in examples as we describe next the DistAlgo language used.

\mypar{DistAlgo, a language for distributed algorithms} 
DistAlgo supports easy distributed programming by building on an
object-oriented programming language, with a formal operational
semantics~\cite{Liu+17DistPL-TOPLAS}, and with an open-source
implementation~\cite{distalgo20git} that extends Python.

Because the implementation uses the Python parser, it uses Python syntax
such as \co{send(m,\,to=p)} instead of the ideal \co{send m to p} for sending
message \co{m} to process \co{p}.  For the same reason, it uses
\co{from\_} in place of the ideal \co{from} because the latter is
a keyword in Python.  A final quirk is that we indicate a previously bound
variable in patterns with prefix \co{\_} in place of the ideal \co{=} because
\co{\_} is the only symbol allowed besides letters and numbers in
identifiers.
Besides the language constructs explained, commonly used constructs in
Python are used, for no operation (\co{pass}), assignments\notes{}
(\co{\p{v}\,=\,\p{e}}), etc.

\mypar{Distributed processes that can send messages}
A type \co{\p{P}} of distributed processes is defined by \co{class
  \p{P}\,(process):~\p{body}}, e.g., lines 1-10 in
Figure~\ref{fig-polling}.
The body may contain
\begin{itemize}
  \setlength{\itemsep}{0ex}

\item a \co{setup} definition for taking in and setting up the values used
  by a type \co{\p{P}} process, e.g., line~2,

\item a \co{run} definition for running the main control flow of the
  process, e.g., lines 3-8,

\item other helper definitions, e.g., lines 9-10, and

\item \co{receive} definitions for handling received messages, e.g., lines
  15-16.

\end{itemize}
A process can refer to itself as \co{self}. Expression \co{self.\p{attr}}
(or \co{\p{attr}} when there is no ambiguity) refers to the value of
\co{\p{attr}} in the process.
\co{\p{ps} = new(\p{P},\p{args},\p{num})} creates \co{\p{num}} (default to
1) new processes of type \co{\p{P}}, optionally passing in the values of
\co{\p{args}} to \co{setup}, and assigns the set of new processes to \co{\p{ps}},
e.g., lines 21 and 22.  \co{start(\p{ps})} starts \co{run()} of processes
\co{\p{ps}}, e.g., lines 23 and 24. A separate \co{setup(\p{ps},\p{args})}
can also set up processes \co{\p{ps}} with the values of \co{\p{args}}.

Processes can send messages: \co{send(\p{m},\,to=\p{ps})} sends message
\co{\p{m}} to processes \co{\p{ps}}, e.g., line~5.

\mypar{Control flow for handling received messages}
Received messages can be handled both asynchronously, using 
\co{receive} definitions, and synchronously, using \co{await}
statements.
\begin{itemize}
  \setlength{\itemsep}{0ex}

\item A \co{receive} definition, \co{def receive (msg=\p{m},
    from\_=\p{p})}, handles, at yield points, un-handled messages that
  match \co{\p{m}} from \co{\p{p}}, e.g., lines 15-16.
  There is a yield point before each \co{await} statement, e.g., line~6,
  for handling messages while waiting.
 The \co{from\_} clause\todo{} is optional.

\item An  \co{await} statement, \co{await} \p{cond}, waits for \p{cond} to be true, e.g.,
  line 6.  A more general statement, \co{if await \p{cond_1}:\,\p{stmt_1}
    elif\,...\,elif \p{cond_k}:\,\p{stmt_k}} \co{elif
    timeout(\p{t}):\,\p{stmt}}, waits for one of \co{\p{cond_1}}, ...,
  \co{\p{cond_k}} to be true or a timeout after \co{\p{t}} seconds, and then
  nondeterministically selects one of \co{\p{stmt_1}}, ...,
  \co{\p{stmt_k}}, \co{\p{stmt}} whose conditions are true to execute.
\end{itemize}\Vex{-1}

\mypar{High-level queries for synchronization conditions}
High-level queries can be used over message histories, and patterns can be
used for matching messages.
\begin{itemize}
  \setlength{\parskip}{.5ex}

\item Histories of messages sent and received by a process are automatically kept in
  variables \co{sent} and \co{received}, respectively.
  \co{sent} is updated at each \co{send} statement, by adding each message
  sent.
  \co{received} is updated at yield points, 
  by adding un-handled messages before executing all matching
  \co{receive} definitions.

  Expression \co{sent(\m{m},\,to=\p{p})} is equivalent to \co{(\p{m},\p{p})
    in sent}.  It returns true iff a message that matches
  \co{(\p{m},\p{p})} is in \co{sent}.
  The \co{to} clause\todo{} is optional.
  \co{received(\m{m},\,from\_=\p{p})} is similar.

\item A pattern can be used to match a message, in \co{sent} and
  \co{received}, and by a \co{receive} definition.  A constant value, such
  as \co{'respond'}, or a previously bound variable, indicated with prefix
  \co{\_}, in the pattern must match the corresponding components of the
  message.  An underscore \co{\_} by itself matches anything.  Previously
  unbound variables in the pattern are bound to the corresponding
  components in the matched message.

  For example, \co{received(('reply','Y',\_t), from\_=r)} on line 7 matches
  in \co{received} every message that is a 3-tuple with components \co{'reply'},
  \co{'Y'}, and the value of \co{t}, and binds \co{r} to the sender.

\end{itemize}
A query can be a comprehension, aggregation, or %
quantification over sets or sequences.
\begin{itemize}
  \setlength{\itemsep}{0ex}

\item A comprehension, \co{setof(\p{e}, \p{v\sb{1}} in \p{s\sb{1}}, ...,
    \p{v\sb{k}} in \p{s\sb{k}}, \p{cond})}, where each \co{\p{v_i}} can be a
  pattern, returns the set of values of \co{\p{e}} for all combinations of
  values of variables that satisfy all \co{\p{v_i} in \p{s\sb{i}}} clauses
  and satisfy condition \co{\p{cond}}, e.g., the comprehension on line~7.

\item An aggregation, similar to a comprehension but with an aggregation
  operator such as \co{countof} or \co{maxof}, returns the value of
  applying the aggregation operator to the collection argument, e.g., the
  \co{countof} query on line 10.

\item A universal quantification, \co{each(\p{v\sb{1}} in \p{s\sb{1}}, ...,
    \p{v\sb{k}} in \p{s\sb{k}}} \co{has=\p{cond}}), returns true iff, for all
  combinations of values of variables that satisfy all \co{\p{v\sb{i}} in
    \p{s\sb{i}}} clauses, \co{\p{cond}} holds, e.g., the \co{each} query on
  line 6.

\item An existential quantification, \co{some(\p{v\sb{1}} in \p{s\sb{1}},
    ..., \p{v\sb{k}} in \p{s\sb{k}}} \co{has=\p{cond})}, returns true iff,
  for some combinations of values of variables that satisfy all
  \co{\p{v\sb{i}} in \p{s\sb{i}}} clauses, \co{\p{cond}} holds, e.g., the
  \co{some} query on line 13.
  When the query returns true, all variables in the query are bound to a
  combination of satisfying values, called a witness, e.g., \co{o} on line
  13. %

\end{itemize}

\mypar{Configuration for setting up and running}
Configuration for requirements such as use of logical clocks and reliable
channels can be specified in a \co{main} definition, e.g., lines 19-20.
When Lamport's logical clocks are used, DistAlgo configures sending and
receiving of a message to update the clock value, and defines a function
\co{logical\_clock()} that returns the clock value.
Processes can then be created, setup, and started.  In general, \co{new} can
have an additional argument, specifying remote nodes where the newly created
processes will run; the default is the local node.

\section{Formal specification of safety and liveness}
\label{sec-spec}

When specifying properties about multiple distributed processes, we refer
to the \co{sent} and \co{received} of a process \p{p} as \co{\p{p}.sent}
and \co{\p{p}.received}.  We use ideal syntax in this section in presenting
the safety and liveness properties, e.g., \co{p.received m from p at t}
instead of \co{p.received(m, from\_=p, clk=t)}.

\mypar{Specifying safety}
Despite being a small and seemingly simple example, a wide variety of
safety properties can be desired for polling.  We consider two of them:
\begin{description}

\item[{\normalfont (S1)}] \emph{The poller has received a reply to the question
    from each pollee when sending the outcome.} %

  This property does not require checking multiple distributed
  processes, because it uses information about only one process, the
  poller.  In fact, in the program in Figure \ref{fig-polling}, it is easy to see that this
  property is implemented clearly in the poller's run method.

  We use this as an example for three reasons: (1) it allows the reader to
  contrast how this is specified and checked by the checker compared with
  by the poller, (2) such checks can be important when we do not have
  access to the internals of a process but can observe messages sent to and
  from the process, and (3) even if we have access to the internals, it may
  be unclear whether the implementation ensures the desired property and
  thus we still need to check it. \todo{}
   
\item[{\normalfont (S2)}] \emph{Each pollee has received the same outcome when
    the program ends.}

  This property requires checking multiple distributed processes, because
  the needed information is not available at a single process.

  We use this example to show such properties can be specified and checked
  easily at the checker, conservatively ensuring safety despite the general
  impossibility results due to message delays, etc.
\end{description}

Consider property (S1). \todo{}The checker will be informed about all
processes (poller \co{p} and pollees \co{rs}) and all messages sent and 
received by each process.  Also, \todo{} it can
use universal and existential quantifications, just as on line 6 of the
poller's code in Figure \ref{fig-polling}. However, there are two issues:
\begin{enumerate}

\item[1)] How does the checker know ``the'' question?
  Inside the poller, ``the'' question is identified by the timestamp in
  variable \co{t}, which is used in the subsequent tests of the replies.
  To check from outside, the checker needs to observe the question and its
  id first, yielding a partial specification for (S1):
\begin{code}
some =p.sent ('question', _, t) has
  each r in rs has some =p.received ('reply', _, =t) from =r
\end{code}
  If one knows that the poller sent only one question, then the \co{some}
  above binds exactly that question.  Otherwise, one could easily check this
  by adding a conjunct\linebreak 
  \co{count \{t:~=p.sent ('question', \_, t)\} == 1}.

\item[2)] How does the checker know that 
  all replies had been received when the outcome was sent
  (Note that a similar question about identifying ``the'' outcome
  can be handled the same way as for ``the'' question.)
  Inside the poller, it is easy to see that tests of the replies occur before the sending of the
  \co{'outcome'} message.
  Outside the poller, we cannot count on the order that the checker
  receives messages to determine the order of events. The checker needs to
  use the timestamps from the logical clocks.
\begin{code}
some =p.sent ('question', _, t), =p.sent ('outcome', _) at t1 has
  each r in rs has some =p.received ('reply', _, =t) from =r at t2 has t1>t2
\end{code}
Note the added \co{=p.sent ('outcome', \_) at t1} on the first line
and \co{at t2 has t1 > t2} on the second line.
\end{enumerate}
Note that when the receiver (respectively sender) or logical time of a sent
(respectively received) message is not used, it is omitted from the
property specification; it could also be matched with an underscore, e.g.,
\co{=p.sent ('question', \_, t) to \_ at \_}.

Consider property (S2), which is now easy, using the same idea to identify
``the'' outcome \co{o} based on the outcome message sent by the poller:
\begin{code}
    some =p.sent ('outcome', o) has
      each r in rs has some =r.received ('outcome', =o)
\end{code}
The checker just needs to check this at the end.

\mypar{Specifying liveness}
Specifying liveness requires language features not used in the algorithm
description.  We use the same specification language we introduced
earlier~\cite{ChaLiu20Liveness-arxiv}.  In particular,
\begin{code}
    evt \p{cond}  
\end{code}
where \co{evt} is read as ``eventually'', denotes that \p{cond} holds 
at some time in the duration that starts from %
the time under discussion,
i.e., eventually, \p{cond} holds.

Many different liveness properties can be desired.  We consider two of them:
\begin{description}

\item[{\normalfont (L1)}] \emph{The poller eventually receives a reply to
    the question.}

  This assumes that a question was sent and covers the duration 
  from then to receiving the first reply.

  We use this example because it is the first indication to the poller that
  the polling really started.  We also use receiving the first reply to
  show a small variation from receiving all replies.

\item[{\normalfont (L2)}] \emph{Eventually each pollee receives the
    outcome.}

  This assumes that an outcome was sent and covers the duration from
  then to all pollees receiving the outcome.

  We use this example because it expresses the completion of the entire
  algorithm.
\end{description}

For (L1), one can simply specify it as the partial specification for (S1)
except with \co{evt} on the second line and with \co{each r in rs} replaced
by \co{some r in rs}:
\begin{code}
    some =p.sent ('question', _, t) has
      evt some r in rs has some =p.received ('reply', _, =t) from =r
\end{code}
In practice, one always estimates an upper bound for message passing time
and poll filling time.  So one can calculate an upper bound on the expected 
time from sending the question to receiving the first reply, and be alerted by 
a timeout if this bound is not met.

For (L2), one can see that this just needs an \co{evt} on the second line
of the specification for (S2):\m{\!}
\begin{code}
    some =p.sent ('outcome', o) has
      evt each r in rs has some =r.received ('outcome', =o)
\end{code}
In practical terms, (L2) implies that the program terminates and (S2) holds
when the program ends.  Thus, with (S2) checked as a safety property, (L2)
mainly checks that the program terminates.

Conceptually, \co{evt} properties are checked against infinite executions.
In practice, they are checked against finite executions by imposing a bound
on when the property should hold, and reporting a violation if the 
property does not hold by then.  From a formal perspective, imposing 
this time bound changes the liveness property to a safety property.

\section{Checking safety}
\label{sec-safe}

We describe a general framework for checking safety through observation by
a checker external to all original processes in the system.  The checker
observes all processes and the messages they send and receive.
We then discuss variations and optimizations.

\mypar{Extending original processes to be checked}  %
The basic idea is: each process \co{p}, when sending or receiving a
message, sends information about the sending or receiving to the
checker.  The checker uses this information to check properties 
of the original processes.

The information sent to the checker may include (1) whether the message is
being sent or received by \co{p}, indicated by \co{'sent'} and \co{'rcvd'},
respectively, (2) the message content, (3) the receiver or receivers (for a
message sent) and sender (for a message received), and (4) the logical
timestamp of the sending or receiving, if a logical clock is used.  In
general, it may include any subset of these, or add any other information
that is available and useful for checking properties of interest.

With ideal channels to send such information to the checker, the checker
can extract all the information using the following correspondence:
\begin{center}\Vex{0.5}
\small
\begin{tabular}{l@{\m{\iff}}l}
  \co{p.sent m to qs at t}  
  & \co{checker received\,('sent' m to qs at t) from p}\\
  \co{p.received m from q at t}
  & \co{checker received\,('rcvd' m from q at t) from p}
\end{tabular}
\end{center}\Vex{.5}

Sending the information can be done by extending the original processes, so
the original program is unchanged.  
The extended processes just need to (1) extend the \co{send} operation to
send information about the sending to the checker, and (2) add a
\co{receive} handler for all messages received to send information about
the receiving to the checker.
A checker process can then specify the safety conditions and check them any
time it desires; to check at the end, it needs to specify a condition to
detect the end.

Figure~\ref{fig-safe} shows safety checking for the polling example.  It
imports the original program\linebreak \co{polling.da} as a module, and extends
processes \co{P} and \co{R} to take \co{checker} as an argument at setup.  In extended
\co{P}, it extends \co{send} and adds \co{receive} to send all 4 kinds of
information listed to \co{checker} (lines 4-8).  In extended \co{R}, it
sends 3 kinds of information, omitting logical times (lines 12-16).
It then defines process \co{Checker} that takes in \co{p} and \co{rs} at
setup, waits for a condition to detect the end of the polling (line 20),
and checks safety properties (S1) and (S2) (lines 22-31).
The \co{main} method is the same as in Figure \ref{fig-polling} except for 
the new and updated lines for adding the checker process, as noted in the comments.

\begin{figure}[ht!]
  \centering
\begin{lstlisting}
import polling  # import the module, polling.da, to be checked 

class P (process, polling.P):  # extend process P in module polling
  def setup(checker,rs): super().setup(rs)  # set up checker and original rs 
  def send(m, to):             # override send to send information to checker
    super().send(m, to)        # do original send
    super().send(('sent', m, to, logical_clock()), to= checker) # send to checker
  def receive(msg= m, from_= fr):  # add a receive matching any message received
    super().send(('rcvd', m, fr, logical_clock()), to= checker) # send to checker
###
#  @timing
#  def polling(): super().polling()
#
#  def run():
#    t = polling()
#    super().send(('time',t), to= checker)
#
#  in send to test lose of msgs
#    import random; 
#    if random.random() < 0.8: super().send(m,to)
#
# ##
class R (process, polling.R):  # as above but extend process R instead
  def setup(checker): super().setup()  # as above but set up only checker
  def send(m, to):             # as above but not send logical time to checker
    super().send(m, to)
    super().send(('sent', m, to), to= checker)
  def receive(msg= m, from_= fr):  # as above but not send logical time to checker
    super().send(('rcvd', m, fr), to= checker) ###
#
#  @timing
#  def polled(): super().polled()
#
#  def run():
#    super().run()
#    t = polled()
#    super().send(('time',t), to= checker) ##

class Checker (process):
  def setup(p,rs): pass        # pass in processes p and rs
  def run(): 
    await each(r in rs, has= some(received(('rcvd', ('outcome',_), _), from_=_r)))
    output('~~ polling ended. checking safety:', S1(), S2()) ###
#           p_sent_one_Q(), each_rs_rcvd_Q(), each_rs_sent_R(),
#           p_rcvd_re_from_all(),
#           p_rcvd_re_from_all_before_sent_O(), # S1()
#           each_rs_rcvd_some_O(),
#           each_rs_rcvd_O()) #S2()
#
#  def p_sent_one_Q():  # p sent one question
#    return countof((Q,t), received(('sent', ('question',Q,t), _,_), from_=_p))
#
#  def each_rs_rcvd_Q():  # each r in rs received some same question from p
#    return some(received(('sent', ('question',q,t), _,_), from_=_p), has=
#                each(r in rs, has= 
#                     received(('rcvd', ('question',q,t), p), from_=r)))
#
#  def each_rs_sent_R():  # each r in rs sent a reply to some same q from p
#    return some(received(('sent', ('question',_,t), _,_), from_=_p), has=
#                each(r in rs, has= 
#                     some(received(('sent', ('reply',_,_t), _p), from_=_r))))
#
#  def p_rcvd_re_from_all():  # p rcvd a re to a q of its from each r in rs 
#    return some(received(('sent', ('question',_,t), _,_), from_=_p), has=
#                each(r in rs, has=
#                     some(received(('rcvd', ('reply',_,_t), _r,_), from_=_p))))
#
#  def each_rs_rcvd_some_O():  # each rs received some outcome
#    return each(r in rs, has= 
#                some(received(('rcvd', ('outcome',_), _), from_=_r)))
# ##
  def S1(): ###
#  def p_rcvd_re_from_all_before_sent_O():
    #some p.sent ('question', _, t), p.sent ('outcome', _) at t1 has
    #  each r in rs has some =p.received('reply',_,=t) from =r at t2 has t1>t2 ##
    return some(received(('sent', ('question',_,t), _, _), from_=_p), 
                received(('sent', ('outcome',_), _, t1), from_=_p), has=
                each(r in rs, has=
                     some(received(('rcvd', ('reply',_,_t), _r, t2), from_=_p),
                          has= t1 > t2)))
  def S2(): ###
#  def each_rs_rcvd_O():  # each rs received the outcome
    #some p.sent ('outcome', o) has
    #  each r in rs has some =r.received ('outcome', =o) from =r ##
    return some(received(('sent', ('outcome',o), _, _), from_=_p), has=
                each(r in rs, has= 
                     some(received(('rcvd', ('outcome',_o), _), from_=_r)))) ###
##
def main():
  config(clock = lamport)
  config(channel = reliable)
  checker = new(Checker)           # create checker
  rs = new(R, [checker], num= 10)  # as in polling but add checker
  p = new(P, [checker, rs])        # as in polling but add checker
  setup(checker, [p,rs])           # setup checker with p and rs
  start(checker)                   # start checker
  start(rs)
  start(p)
\end{lstlisting}
  \caption{Checking safety for the polling program.}
  \label{fig-safe}
\end{figure}

\mypar{Variations and optimizations}\todo{}
When checking systems with many processes, a single checker process would
be a bottleneck.  The single checker framework can easily be extended to
use a hierarchy of checkers, in which each checker observes a subset of
original processes and/or child checkers and reports to a parent checker.

As an optimization, information not needed for the properties being checked can be
omitted from messages sent to the checker, leading to more efficient executions 
and simpler patterns in
specifying the conditions to be checked.  In Figure~\ref{fig-safe}, process
\co{R} already omits logical times from all messages to the
checker.  More information can be omitted.  For example, process \co{P} can
omit target processes, the 3rd component of the message, in information about sending.
Additionally, process \co{R} can omit all information about sending and all
but the second component about receiving.
Furthermore, more refined patterns can be used when extending \co{send} to
omit unused parts inside the message content, the second component.  For
example, the specific question in \co{'question'} messages can be omitted.

Instead of extending the original processes to be checked, an alternative
is to let the original processes extend a checker process.  While the
former approach requires no change at all to the original processes, the latter
approach requires small changes: (1) to each original process class, (a) add the
checker process class as a base class and (b) add a \co{setup} parameter to
pass in the checker process, and (2) in \co{main}, (a) create, setup, and
start the checker process and (b) in the call to \co{new} or \co{setup} for
each original process, add the checker process as an additional argument.
The advantage of this alternative approach is that the same checker class can be
used for checking different programs when the same checking is desired.  An
example use is for benchmarking the \co{run} method of different
programs\footnote{\url{http://github.com/DistAlgo/distalgo/blob/master/benchmarks/controller.da}}.

While checking safety using our framework is already relatively easy,
higher-level declarative languages can be designed for specifying the
desired checks, and specifications in such languages can be compiled into
optimized checking programs that require no manual changes to the original
programs.

\section{Checking liveness}
\label{sec-live}

As discussed in Section~\ref{sec-spec}, in finite executions, checking
liveness boils down to safety checking plus use of timeouts to check that
properties hold within an expected amount of time.
For the polling example, checking timeouts plus the same or similar conditions as
(S1) and (S2) corresponds to what can be checked for (L1) and
(L2).  We describe a general framework for easily checking
timeouts during program execution based on elapsed real time at the checker process.   
Using real time at the checker avoids assumptions about clock synchronization.
We then discuss variations and optimizations.

\mypar{Extending original processes to be checked} 
The same framework to extend original processes for safety checking can be
used for liveness checking.  One only needs to specify checks for timeouts instead of or in
addition to safety checks.  We show how timeouts between observations of
any two sending or receiving events, as well as a timeout for the entire
execution, can easily be checked, even with multiple timeout checks running concurrently.

Given any two sending or receiving events \co{e1} and \co{e2}, we check
that after \co{e1} is observed by the checker, \co{e2} is observed within 
a specified time bound.  There are two steps:
\begin{itemize}

\item[1)] When the checker receives \co{e1}, it starts a timer for the
  specified time bound.  Each timer runs in a
  separate thread and, when it times out, it sends a timeout message to
  the checker.

\item[2)] When the checker receives a timeout message, it checks whether
  the expected event \co{e2} has been observed.  If yes, it does nothing.  Otherwise, it
  reports a violation of the time bound requirement.

\end{itemize}
All time bounds are specified in a map, which maps a name for a pair of
events to the required time bound from observing the first event until
observing the second event.

This framework can easily be generalized to check conditions involving any
number of events.  When the timeout happens, instead of checking whether
one specific event \co{e2} has been observed, the checker can check whether
several expected events have all been observed, or whether any other
desired condition on the set of observed events holds.
 
A time bound for the entire execution of the algorithm can be set and
checked separately, in addition to checking any other safety and liveness
properties, to directly check overall termination, using an appropriate
condition to detect whether the algorithm has completed successfully.

Figure~\ref{fig-live} shows timeout checking for the polling example.  To
check liveness instead of safety, one could use exactly the same program as
for safety check except for the added \co{import}'s and \co{TIMEOUTS} map at
the top and a rewritten \co{Checker} process.  To check timeouts in
addition to safety, the \co{Checker} process in Figure \ref{fig-live} can extend the
\co{Checker} process in Figure~\ref{fig-safe}, and just add the function calls
\co{super().S1()} and \co{super().S2()} at the end of the \co{run} method here.

Modules \co{threading} and \co{time} are imported in order to run each timer in
a separate thread.  A dictionary \co{TIMEOUTS} holds the map of time bounds
(in seconds) for different pairs of events: \co{'q-r'} is for the poller sending
the question and the poller receiving the first reply, corresponding to (L1),
and \co{'o-o'} is for the poller sending the outcome and all pollees receiving the
outcome, corresponding to (L2).  The dictionary also includes an entry \co{'total'}
with a time bound for the entire execution of the algorithm.

The \co{Checker} process waits for the same condition, as for safety checking, %
to detect the end of the polling (line 8-9), but with a timeout 
for \co{'total'} (line 10) while waiting.  It starts two
timers corresponding to (L1) and (L2) when observing the question was sent
(lines 13-16), and checks and reports timeouts when any timer times out (lines
23-30).

\begin{figure}[t!]
  \centering
\begin{lstlisting}
import threading
import time ###
#import sys
from polling_check import P,R ##
TIMEOUTS = {'q-r':0.00001, 'o-o':0.0001, 'total':1}
                                      # map of timeout names to time bounds
class Checker (process):
  def setup(p,rs): pass
  def run(): ###
#    while True:
#      if await some(received(('sent', ('question',_,t), p, _))): # repeated:(
#        timer(('q-r',t,p))
#        timer(('q-o',t,p)) ##
    if await each(r in rs, has= 
                  some(received(('rcvd', ('outcome',_), _), from_=_r))): pass
    elif timeout(TIMEOUTS['total']):  # entire execution times out
      output('!! total timeout receiving outcome by all pollees') ###
#   await timeout(TIMEOUTS['total'])  # entire execution times out 
#   output('~~ polling ended. checking safety:', S1(), S2()) ##
    output('~~ polling ended. timeouts checked')

  def receive(msg=('sent', ('question',_,t), ps, _)):  # at 1st of two events
    runtimer('q-r',t,ps)  # start timer 'q-r' that starts at this event
  def receive(msg=('sent', ('outcome',o), ps, _)):  # at 1st of two events
    runtimer('o-o',o,ps)  # start timer 'o-o' that starts at this event
###
#  import functools
#  @functools.lru_cache(maxsize=None) # memo to not runtimer for save event ## 
  def runtimer(*args):   # run timer in a separate thread passing in all args
    threading.Thread(target=timer, args=args).start() ###
# from sending reply to receiving outcome.
# from sending a question, to receiving the question.
##
  def timer(name, *rest):  # timer, taking timer name and rest of arguments
    time.sleep(TIMEOUTS[name])  # sleep the time bound for the timer name
    send(('timeout', name, *rest), to= self)  # send timeout to checker itself

  def receive(msg=('timeout', 'q-r', t, ps)):  # at timeout for 'q-r'
    if not some(r in rs, has=  # check 2nd event
                some(received(('rcvd', ('reply',_,_t), _, _), from_=_p))):
      output('!! L1 timeout receiving the first reply by the poller', t, ps) ###
##
  def receive(msg=('timeout', 'o-o', o, ps)):  # at timeout for 'o-o'
    if not each(r in rs, has=  # check 2nd event
                some(received(('rcvd', ('outcome',_o), _), from_=_r))): 
      output('!! L2 timeout receiving outcome by all pollees', o, ps, r) ###
#      sys.exit()

def main():
  config(clock = lamport)
  config(channel = reliable)
  checker = new(Checker)           # create checker
  rs = new(R, [checker], num= 10)  # as in polling but add checker
  p = new(P, [checker, rs])        # as in polling but add checker
  setup(checker, [p,rs])           # setup checker with p and rs
  start(checker)                   # start checker
  start(rs)
  start(p) ##
\end{lstlisting}
  \caption{Checking timeouts for the polling program.}
  \label{fig-live}
\end{figure}

\mypar{Variations and optimizations}
Variations and optimizations for checking safety can also be used for
checking liveness.  Checking timeouts using real time is the additional
challenge.

In the program in Figure~\ref{fig-live}, the timeout \co{'o-o'} for (L2)
waits for essentially the same condition as for detecting the end of
polling in \co{run}.
Therefore, the test for
detecting the end of polling in \co{run} is unnecessary in this case, and
the entire body of \co{run} may simply be \co{await timeout(TIMEOUTS['total'])}.
When a timeout \co{'o-o'} is received, the checker could terminate itself
by importing \co{sys} and calling \co{sys.exit()}.  Of course after either
timeout for \co{'o-o'} or timeout in \co{run}, the checker process could
also do anything else helpful instead of terminating itself.

Instead of or in addition to using real time\todo{} at the checker, one
could use real time at the original processes.  A clock synchronization 
algorithm can be used to improve the precision of clocks at different 
original processes and the overall precision and time bounds.
Even without clock synchronization, using real time at the original processes
can improve the precision and bounds for liveness properties involving 
multiple events at the same process, such as (L1).

Note that observing the start and end of an operation or entire program is
also how performance measurements can be performed, as mentioned for
benchmarking at the end of Section~\ref{sec-safe}.

\section{Implementation and experiments}
\label{sec-impl}

DistAlgo has been implemented as an extension of the Python programming
language and used extensively in studying and teaching of distributed
algorithms~\cite{Liu+17DistPL-TOPLAS}.
The framework discussed for checking safety and liveness properties has
also been used extensively, in both ad hoc and more systematic ways.
We describe using the DistAlgo implementation and our framework for
execution and runtime checking.

\mypar{Execution and fault simulation}
DistAlgo is open source and available on github~\cite{distalgo20git}.  One can simply
add it to the Python path after downloading, and run the \co{da} module in
Python, e.g., running the program \co{polling.da} in
Figure~\ref{fig-polling} as \co{python -m da polling.da}.

For runtime checking, a checking program, such as the program in
Figure~\ref{fig-safe} can be run in the same way.  More generally,
implementations of three conceptual components are needed:
\begin{itemize}

\item[1)] A distributed algorithm, plus input taken by the algorithm.  This
  needs a complete executable program, such as \co{polling.da} in
  Figure~\ref{fig-polling}.

\item[2)] Safety and liveness requirements to be checked.
  These are expressed as executable functions that can be called at required points
  during the execution, such as functions \co{S1} and \co{S2} in
  Figure~\ref{fig-safe} and the \co{receive} handlers in Figure~\ref{fig-live}.

\item[3)] Process and communication failures to be considered.  These
  correspond to executable configurations that can be executed for fault
  simulation, as described below.

\end{itemize}
Our framework puts these together naturally, with small configurations to
observe processes and communications, with both logical and real times,
thanks to the power of the DistAlgo language and compiler.

Fault simulation is essential for checking safety and liveness of complex
algorithms in simulation of real distributed systems that are fault-prone.
Both process and communication failures may happen, but the latter are much
more frequent.  Also, the former can be simulated with the latter, because
a process interacts with other processes only through communication.  For
example, a crashed process is indistinguishable from one that stops sending 
messages to other processes.

With our existing framework for checking, we can simply use \co{send} to
simulate all possible communication failures, including message
\begin{itemize}
\item loss: drop messages without sending;
\item delay: delay messages before sending; 
\item duplication: send messages multiple times;
\item reordering: delay sending messages until after sending later messages; and 
\item corruption for simulating Byzantine failures: change message before sending.
\end{itemize}

For example, to simulate a 1\% chance of dropping a message sent by a
\co{P} process, in the \co{send} method in Figure~\ref{fig-safe}, we can
put \co{super().send(m,to)} inside a conditional:
\begin{code}
  if random.random() < 0.99: super().send(m,to)
\end{code}
and add \co{import random} before it.

Similarly, one may add fixed or random delays, duplications, reorderings,
or a combination of them.  A main issue to note is
that, in general, one would want to send a delayed or duplicated message
using a separate thread, to avoid delaying the execution of the algorithm.

Configuration options and values can be provided through command-line
arguments and external files, as well as built-in language constructs.  All
these kinds have been provided in the DistAlgo language and compiler and
used in DistAlgo programs.  Similar mechanisms have been used in all kinds
of system configurations for decades.  A challenge
is to design and implement a powerful, commonly accepted language for such
configurations.

\mypar{Experiments and experience}
For the running example, checking both safety and liveness, with the
\co{Checker} process in Figure~\ref{fig-live} extending that in
Figure~\ref{fig-safe} but replacing the last line in \co{run} in
Figure~\ref{fig-live} with
the last line in \co{run} in Figure~\ref{fig-safe} and adding
\co{super().}\,before \co{S1()} and \co{S2()}, an example output is
as shown below:

{\footnotesize
\begin{verbatim}
> python -m da .\polling_check_live.da
[56] da.api<MainProcess>:INFO: <Node_:e8001> initialized at 127.0.0.1:(UdpTransport=56264, T
cpTransport=16135).
[56] da.api<MainProcess>:INFO: Starting program <module 'polling_check_live' from '.\\pollin
g_check_live.da'>...
[58] da.api<MainProcess>:INFO: Running iteration 1 ...
[62] da.api<MainProcess>:INFO: Waiting for remaining child processes to terminate...(Press "
Ctrl-Brk" to force kill)
[1748] da.api<MainProcess>:INFO: Main process terminated.
[144] polling_check.P<P:bb00d>:OUTPUT: -- received Y from: {<R:bb003>, <R:bb007>, <R:bb00b>,
 <R:bb009>, <R:bb004>, <R:bb008>}
[568] polling_check.R<R:bb00a>:OUTPUT: == received outcome: 6
[1526] polling_check.R<R:bb003>:OUTPUT: == received outcome: 6
[975] polling_check.R<R:bb007>:OUTPUT: == received outcome: 6
[1259] polling_check.R<R:bb005>:OUTPUT: == received outcome: 6
[436] polling_check.R<R:bb00b>:OUTPUT: == received outcome: 6
[304] polling_check.R<R:bb00c>:OUTPUT: == received outcome: 6
[1391] polling_check.R<R:bb004>:OUTPUT: == received outcome: 6
[710] polling_check.R<R:bb009>:OUTPUT: == received outcome: 6
[1121] polling_check.R<R:bb006>:OUTPUT: == received outcome: 6
[1671] polling_check_live.Checker<Checker:bb002>:OUTPUT: !! L2 timeout receiving outcome by 
all pollees 6 {<R:bb007>, <R:bb009>, <R:bb004>, <R:bb006>, <R:bb00a>, <R:bb005>, <R:bb00b>, 
<R:bb00c>, <R:bb003>, <R:bb008>} <R:bb008>
[852] polling_check.R<R:bb008>:OUTPUT: == received outcome: 6
[1676] polling_check_live.Checker<Checker:bb002>:OUTPUT: ~~ polling ended. checking safety: 
True True
\end{verbatim}}

\noindent 
Notice the last process, \co{<R:bb008>}, printed at the end of the 3 
lines reporting (L2) timeout; it shows a witness for violation of the
\co{each} check on lines 28-29 in Figure~\ref{fig-live}, printed at the end
of line 30.
When we increased the timeout for \co{'o-o'} to 0.01 seconds, no (L2)
timeout was reported in all dozens of runs checked.
When we added message loss rate of 10\%, we saw some runs reporting total
timeout, and some runs reporting even all three timeouts.

Overall, we have used the checking framework in implementation, testing,
debugging, simulation, and analysis of many well-known distributed
algorithms, and in developing their high-level executable specifications.
This includes a variety of algorithms for distributed mutual exclusion and
distributed consensus written in
DistAlgo~\cite{Liu+12DistSpec-SSS,Liu+12DistPL-OOPSLA,Liu18LclockUnfair-APPLIED,Liu+19Paxos-PPDP},
especially including over a dozen well-known algorithms and variants for
classical consensus and blockchain
consensus~\cite{LiuSto19DistConsensus-PODCtut}.
Use of DistAlgo has helped us find improvements to both correctness and
efficiency of well-known distributed algorithms, 
e.g.,~\cite{Liu+12DistSpec-SSS,Liu18LclockUnfair-APPLIED,Liu+19Paxos-PPDP}.

We have also used the framework in other research,
e.g.,~\cite{Liu+16IncOQ-PPDP}, and in teaching distributed algorithms and
distributed systems to help study and implement many more algorithms.
DistAlgo has been used by hundreds of students in graduate and
undergraduate courses in over 100 different course projects, implementing
and checking the core of network protocols, %
distributed graph algorithms, distributed coordination
services, %
distributed hash tables, %
distributed file systems, %
distributed databases, %
parallel %
processing platforms, %
security protocols, and more~\cite{Liu+17DistPL-TOPLAS}.

The algorithms and systems can be programmed much more easily and clearly
compared to using conventional programming languages, e.g., in 20 lines
instead of 200 lines,
or 300 lines instead of 3000 lines or many more.  Systematic methods for
checking these algorithms and implementations has been a continual effort.

Additional information is available at
\url{http://distalgo.cs.stonybrook.edu/tutorial}.

\section{Related work} %
\label{sec-related}

Francalanza et al. broadly surveyed runtime verification research related to distributed systems \cite{Francalanza2018}.  Here, we focus on aspects related to DistAlgo.

\mypar{Global property detection}
Many algorithms have been developed to detect global properties in distributed systems, e.g.,
\cite{Garg02,KshSin08}.  These algorithms vary along several dimensions.  
For example, many consider only the happened-before ordering \cite{Lam78}; 
others also exploit orderings from approximately-synchronized real-time
clocks~\cite{stoller2000detecting}.  Some can detect arbitrary predicates; 
other are specialized to check a class of properties efficiently.  
Many use a single checker process (as in our example); 
others use a hierarchy of checker processes, or are decentralized, 
with the locus of control moving among the monitored processes.  
DistAlgo's high-level nature makes it very well-suited for specifying 
and implementing all such algorithms.

\mypar{Efficient invariant checking}
Runtime checking of invariants, in centralized or distributed systems, requires evaluating them repeatedly.  This can be expensive for complex invariants, especially invariants that involve nested quantifiers.  We used incrementalization for efficient repeated evaluation of predicates in the contexts of runtime invariant checking and query-based debugging for Python programs \cite{Gor+08RTInvCheck-WODA,Gor+08QBDebug-SCAM}.  We later extended our incrementalization techniques to handle quantifications in DistAlgo programs \cite{Liu+17DistPL-TOPLAS}.

\mypar{Centralization}
Due to the difficulty of runtime checking of truly distributed systems, some approaches create centralized versions of them.  We have developed a source-level centralization transformation for DistAlgo that produces a non-deterministic sequential program, well-suited to simulation and verification.  In prior work, we developed bytecode-level transformations that transform a distributed Java program using Remote Method Invocation (RMI) into a centralized Java program using simulated RMI \cite{StoLiu01TransMC-SPIN}.
Minha \cite{machado2019} takes another approach to centralization of distributed Java programs, by virtualizing multiple Java Virtual Machine (JVM) instances in a single JVM and providing a library that simulates network communication.

\mypar{DistAlgo translators}
Grall et al. developed an automatic translation from Event-B models of distributed algorithms to DistAlgo \cite{grall2020}.  Event-B is a modeling language adapted to verification of distributed algorithms.  They chose DistAlgo as the target language because ``Its high-levelness makes DistAlgo closer to the mathematical notations of Event-B and improves the clarity of DistAlgo programs.''  We developed a translator from DistAlgo to TLA+, allowing verification tools for TLA+ to be applied to the translations \cite{Liu+8DistInv-TLA}.

\mypar{Conclusion}
We have presented a general, simple, and complete framework for runtime
checking of distributed algorithms.  The framework, as well as the
algorithms and properties to be checked, are written in a high-level
language that is both completely precise and directly executable.  A
challenging problem for future work is a powerful, commonly accepted
language for higher-level, declarative configurations for checking
distributed algorithms and systems.

\def\bibdir{../../../../bib}      %
\renewcommand{\baselinestretch}{-.9}
\small%
\bibliography{\bibdir/strings,\bibdir/liu,\bibdir/IC,\bibdir/PT,\bibdir/Lang,\bibdir/Algo,\bibdir/DB,\bibdir/AI,\bibdir/Sec,\bibdir/Sys,\bibdir/Veri,\bibdir/misc,\bibdir/stoller,\bibdir/crossref,new}

\newcommand{\etalchar}[1]{$^{#1}$}
\begin{thebibliography}{MMN{\etalchar{+}}19}

\bibitem[BJ87a]{birman1987vs}
K.~Birman and T.~Joseph.
\newblock Exploiting virtual synchrony in distributed systems.
\newblock In {\em Proceedings of the 11th ACM Symposium on Operating Systems
  Principles}, pages 123--138. ACM Press, Nov. 1987.

\bibitem[BJ87b]{birman1987reliable}
Kenneth~P Birman and Thomas~A Joseph.
\newblock Reliable communication in the presence of failures.
\newblock {\em ACM Transactions on Computer Systems (TOCS)}, 5(1):47--76, 1987.

\bibitem[BMR10]{Bir+10virtualsync}
Ken Birman, Dahlia Malkhi, and Robbert~Van Renesse.
\newblock Virtually synchronous methodology for dynamic service replication.
\newblock Technical Report MSR-TR-2010-151, Microsoft Research, 2010.

\bibitem[CL20]{ChaLiu20Liveness-arxiv}
Saksham Chand and Yanhong~A. Liu.
\newblock What's live? understanding distributed consensus.
\newblock {\em Computing Research Repository}, \normalfont {arXiv:2001.04787
  [cs.DC]}, Jan. 2020.
\newblock \url{http://arxiv.org/abs/2001.04787}.

\bibitem[CLS16]{Cha+16PaxosTLAPS-FM}
Saksham Chand, Yanhong~A. Liu, and Scott~D. Stoller.
\newblock Formal verification of {Multi-Paxos} for distributed consensus.
\newblock In {\em Proceedings of the 21st International Symposium on Formal
  Methods}, volume 9995 of {\em Lecture Notes in Computer Science}, pages
  119--136. Springer, Nov. 2016.

\bibitem[FLP85]{fischer85flp}
Michael~J. Fischer, Nancy~A. Lynch, and Michael~S. Paterson.
\newblock Impossibility of distributed consensus with one faulty process.
\newblock {\em Journal of the ACM}, 32(2):374--382, Apr. 1985.

\bibitem[Fok13]{Fok13}
Wan Fokkink.
\newblock {\em Distributed Algorithms: An Intuitive Approach}.
\newblock MIT Press, 2013.

\bibitem[FPS18]{Francalanza2018}
Adrian Francalanza, Jorge~A. P{\'e}rez, and C{\'e}sar S{\'a}nchez.
\newblock {\em Runtime Verification for Decentralised and Distributed Systems},
  volume 10457 of {\em Lecture Notes in Computer Science}, pages 176--210.
\newblock Springer, 2018.

\bibitem[Gar02]{Garg02}
Vijay~K. Garg.
\newblock {\em Elements of Distributed Computing}.
\newblock Wiley, 2002.

\bibitem[Gra20]{grall2020}
Alexis Grall.
\newblock {Automatic Generation of DistAlgo Programs from Event-B Models}.
\newblock In {\em International Conference on Rigorous State-Based Methods
  ({ABZ} 2020)}, pages 414--417. Springer, 2020.

\bibitem[GRLS08]{Gor+08RTInvCheck-WODA}
Michael Gorbovitski, Tom Rothamel, Yanhong~A. Liu, and Scott~D. Stoller.
\newblock Efficient runtime invariant checking: {A} framework and case study.
\newblock In {\em Proceedings of the 6th International Workshop on Dynamic
  Analysis}, pages 43--49. ACM Press, 2008.

\bibitem[GTR{\etalchar{+}}08]{Gor+08QBDebug-SCAM}
Michael Gorbovitski, K.~Tuncay Tekle, Tom Rothamel, Scott~D. Stoller, and
  Yanhong~A. Liu.
\newblock Analysis and transformations for efficient query-based debugging.
\newblock In {\em Proceedings of the 8th IEEE International Working Conference
  on Source Code Analysis and Manipulation}, pages 174--183. IEEE CS Press,
  2008.

\bibitem[HHK{\etalchar{+}}15]{hawblitzel2015ironfleet}
Chris Hawblitzel, Jon Howell, Manos Kapritsos, Jacob~R. Lorch, Bryan Parno,
  Michael~L. Roberts, Srinath Setty, and Brian Zill.
\newblock {IronFleet}: {P}roving practical distributed systems correct.
\newblock In {\em Proceedings of the 25th Symposium on Operating Systems
  Principles}, pages 1--17. ACM Press, 2015.

\bibitem[KLC{\etalchar{+}}19]{Kan+18CryptoAbs-PLAS}
Christopher Kane, Bo~Lin, Saksham Chand, Scott~D. Stoller, and Yanhong~A. Liu.
\newblock High-level cryptographic abstractions.
\newblock In {\em Proceedings of the ACM SIGSAC 14th Workshop on Programming
  Languages and Analysis for Security}, London, U.K., Nov. 2019. ACM Press.

\bibitem[KS08]{KshSin08}
A.D. Kshemkalyani and M.~Singhal.
\newblock {\em Distributed Computing: Principles, Algorithms, and Systems}.
\newblock Cambridge University Press, 2008.

\bibitem[Lam77]{lamport1977proving}
Leslie Lamport.
\newblock Proving the correctness of multiprocess programs.
\newblock {\em IEEE transactions on software engineering}, 3(2):125--143, 1977.

\bibitem[Lam78]{Lam78}
Leslie Lamport.
\newblock Time, clocks, and the ordering of events in a distributed system.
\newblock {\em Communications of the ACM}, 21(7):558--565, 1978.

\bibitem[Lam94]{lam94tla}
Leslie Lamport.
\newblock The temporal logic of actions.
\newblock {\em ACM Transactions on Programming Languages and Systems},
  16(3):872--923, 1994.

\bibitem[Lam98]{Lam98paxos}
Leslie Lamport.
\newblock The part-time parliament.
\newblock {\em ACM Transactions on Computer Systems}, 16(2):133--169, 1998.

\bibitem[Lam02]{lam02book}
Leslie Lamport.
\newblock {\em Specifying Systems: The TLA+ Language and Tools for Hardware and
  Software Engineers}.
\newblock Addison-Wesley, 2002.

\bibitem[LBSL16]{Liu+16IncOQ-PPDP}
Yanhong~A. Liu, Jon Brandvein, Scott~D. Stoller, and Bo~Lin.
\newblock Demand-driven incremental object queries.
\newblock In {\em Proceedings of the 18th International Symposium on Principles
  and Practice of Declarative Programming}, pages 228--241. ACM Press, 2016.

\bibitem[LC12]{liskov2012vr}
Barbara Liskov and James Cowling.
\newblock Viewstamped replication revisited.
\newblock Computer Science and Artificial Intelligence Laboratory Technical
  Report MIT-CSAIL-TR-2012-021, Massachusetts Institute of Technology,
  Cambridge, MA 02139, USA, 2012.

\bibitem[LCS19]{Liu+19Paxos-PPDP}
Yanhong~A. Liu, Saksham Chand, and Scott~D. Stoller.
\newblock Moderately complex {Paxos} made simple: High-level executable
  specification of distributed algorithm.
\newblock In {\em Proceedings of the 21st International Symposium on Principles
  and Practice of Declarative Programming}, pages 15:1--15:15. ACM Press, Oct.
  2019.

\bibitem[Liu18]{Liu18LclockUnfair-APPLIED}
Yanhong~A. Liu.
\newblock Logical clocks are not fair: {W}hat is fair? {A} case study of
  high-level language and optimization.
\newblock In {\em Proceedings of the Workshop on Advanced Tools, Programming
  Languages, and Platforms for Implementing and Evaluating Algorithms for
  Distributed Systems}, pages 21--27. ACM Press, 2018.

\bibitem[LL20]{distalgo20git}
Bo~Lin and Yanhong~A. Liu.
\newblock {DistAlgo}: A language for distributed algorithms.
\newblock \url{http://github.com/DistAlgo}, 2014 (Last release March 2020).

\bibitem[LS19]{LiuSto19DistConsensus-PODCtut}
Yanhong~A. Liu and Scott~D. Stoller.
\newblock From classical to blockchain consensus: {What} are the exact
  algorithms?
\newblock In {\em Proceedings of the 2019 ACM Symposium on Principles of
  Distributed Computing}, pages 544--545. ACM Press, July-Aug. 2019.

\bibitem[LSCW18]{Liu+8DistInv-TLA}
Yanhong~A. Liu, Scott~D. Stoller, Saksham Chand, and Xuetian Weng.
\newblock Invariants in distributed algorithms.
\newblock In {\em Proceedings of the TLA+ Community Meeting}, Oxford, U.K.,
  2018.
\newblock \url{http://www.cs.stonybrook.edu/~liu/papers/DistInv-TLA18.pdf}.

\bibitem[LSL12]{Liu+12DistSpec-SSS}
Yanhong~A. Liu, Scott~D. Stoller, and Bo~Lin.
\newblock High-level executable specifications of distributed algorithms.
\newblock In {\em Proceedings of the 14th International Symposium on
  Stabilization, Safety, and Security of Distributed Systems}, pages 95--110.
  Springer, 2012.

\bibitem[LSL17]{Liu+17DistPL-TOPLAS}
Yanhong~A. Liu, Scott~D. Stoller, and Bo~Lin.
\newblock From clarity to efficiency for distributed algorithms.
\newblock {\em ACM Transactions on Programming Languages and Systems},
  39(3):12:1--12:41, May 2017.

\bibitem[LSLG12]{Liu+12DistPL-OOPSLA}
Yanhong~A. Liu, Scott~D. Stoller, Bo~Lin, and Michael Gorbovitski.
\newblock From clarity to efficiency for distributed algorithms.
\newblock In {\em Proceedings of the 27th ACM SIGPLAN Conference on
  Object-Oriented Programming, Systems, Languages and Applications}, pages
  395--410. ACM Press, 2012.

\bibitem[Lyn96]{Lynch96}
Nancy~A. Lynch.
\newblock {\em Distributed Algorithms}.
\newblock Morgan Kaufman, 1996.

\bibitem[{Mic}20]{tlatoolbox20}
{Microsoft Research}.
\newblock {The TLA Toolbox}.
\newblock \url{http://lamport.azurewebsites.net/tla/toolbox.html}, Last
  modified April 27, 2020.

\bibitem[MMN{\etalchar{+}}19]{machado2019}
Nuno Machado, Francisco Maia, Francisco Neves, F{\'{a}}bio Coelho, and
  Jos{\'{e}} Pereira.
\newblock Minha: Large-scale distributed systems testing made practical.
\newblock In Pascal Felber, Roy Friedman, Seth Gilbert, and Avery Miller,
  editors, {\em 23rd International Conference on Principles of Distributed
  Systems ({OPODIS} 2019)}, volume 153 of {\em LIPIcs}, pages 11:1--11:17.
  Schloss Dagstuhl - Leibniz-Zentrum f{\"{u}}r Informatik, 2019.

\bibitem[MP20]{mcmillan2020ivy}
Kenneth~L McMillan and Oded Padon.
\newblock Ivy: A multi-modal verification tool for distributed algorithms.
\newblock In {\em International Conference on Computer Aided Verification},
  pages 190--202. Springer, 2020.

\bibitem[OL88]{oki88vsr}
Brian~M Oki and Barbara~H Liskov.
\newblock Viewstamped replication: {A} new primary copy method to support
  highly-available distributed systems.
\newblock In {\em Proceedings of the 7th Annual ACM Symposium on Principles of
  Distributed Computing}, pages 8--17. ACM Press, 1988.

\bibitem[PLSS17]{padon2017paxos}
Oded Padon, Giuliano Losa, Mooly Sagiv, and Sharon Shoham.
\newblock Paxos made {EPR}: {Decidable} reasoning about distributed protocols.
\newblock {\em Proceedings of the ACM on Programming Languages}, 1(OOPSLA):108,
  2017.

\bibitem[Ray86]{raynal1986algorithms}
Michel Raynal.
\newblock {\em Algorithms for Mutual Exclusion}.
\newblock MIT Press, 1986.

\bibitem[SL01]{StoLiu01TransMC-SPIN}
Scott~D. Stoller and Yanhong~A. Liu.
\newblock Transformations for model checking distributed java programs.
\newblock In {\em Proceedings of the 8th International SPIN Workshop on Model
  Checking of Software}, volume 2057 of {\em Lecture Notes in Computer
  Science}, pages 192--199. Springer, May 2001.

\bibitem[Sto00]{stoller2000detecting}
Scott~D. Stoller.
\newblock Detecting global predicates in distributed systems with clocks.
\newblock {\em Distributed Computing}, 13(2):85--98, Apr. 2000.

\end{thebibliography}
\bibliographystyle{alpha}

\end{document}